\documentclass{bmcart}
\usepackage[utf8]{inputenc} %unicode support
\usepackage{graphicx}% Include figure files
\usepackage{dcolumn}% Align table columns on decimal point
\usepackage{bm}% bold math
\usepackage{amssymb,amsfonts,amsmath}
\usepackage{bbold}
\usepackage[titletoc,title]{appendix}
\startlocaldefs

\endlocaldefs

\begin{document}

%%% Start of article front matter
\begin{frontmatter}

\begin{fmbox}
\dochead{Research}

\title{Numerical calculation of protein-ligand binding rates through solution of the Smoluchowski equation using smooth particle hydrodynamics}

%%%%%%%%%%%%%%%%%%%%%%%%%%%%%%%%%%%%%%%%%%%%%%
%%                                          %%
%% Enter the authors here                   %%
%%                                          %%
%% Specify information, if available,       %%
%% in the form:                             %%
%%   <key>={<id1>,<id2>}                    %%
%%   <key>=                                 %%
%% Comment or delete the keys which are     %%
%% not used. Repeat \author command as much %%
%% as required.                             %%
%%                                          %%
%%%%%%%%%%%%%%%%%%%%%%%%%%%%%%%%%%%%%%%%%%%%%%

\author[
   addressref={wenxiao},
   email={wenxiao.pan@pnnl.gov} 
]{\inits{WP}\fnm{Wenxiao} \snm{Pan}}
\author[
   addressref={mike},
   email={michael.daily@pnnl.gov}
]{\inits{MD}\fnm{Michael} \snm{Daily}}
\author[
	addressref={nathan},
	corref={nathan},
	email={nathan.baker@pnnl.gov}
]{\inits{NAB}\fnm{Nathan A.} \snm{Baker}}

%%%%%%%%%%%%%%%%%%%%%%%%%%%%%%%%%%%%%%%%%%%%%%
%%                                          %%
%% Enter the authors' addresses here        %%
%%                                          %%
%% Repeat \address commands as much as      %%
%% required.                                %%
%%                                          %%
%%%%%%%%%%%%%%%%%%%%%%%%%%%%%%%%%%%%%%%%%%%%%%

\address[id=wenxiao]{% % unique id
  \orgname{Advanced Computing, Mathematics and Data Division, Pacific Northwest National Laboratory}, % university, etc
  \street{PO Box 999, MSID K7-90}, %
  \postcode{99352}, % post or zip code
  \city{Richland, WA}, % city
  \cny{USA} % country
}
\address[id=mike]{% % unique id
  \orgname{Chemical Physics and Analysis Division, Mathematics and Data Division, Pacific Northwest National Laboratory}, % university, etc
  \street{PO Box 999, MSID K1-83}, %
  \postcode{99352}, % post or zip code
  \city{Richland, WA}, % city
  \cny{USA} % country
}

\address[id=nathan]{% % unique id
  \orgname{Computational and Statistical Analytics Division, Pacific Northwest National Laboratory}, % university, etc
  \street{PO Box 999, MSID K7-20}, %
  \postcode{99352}, % post or zip code
  \city{Richland, WA}, % city
  \cny{USA} % country
}

%%%%%%%%%%%%%%%%%%%%%%%%%%%%%%%%%%%%%%%%%%%%%%
%%                                          %%
%% Enter short notes here                   %%
%%                                          %%
%% Short notes will be after addresses      %%
%% on first page.                           %%
%%                                          %%
%%%%%%%%%%%%%%%%%%%%%%%%%%%%%%%%%%%%%%%%%%%%%%

%%\note{Sample of title note}     % note to the article
%\note[id=n1]{Equal contributor} % note, connected to author

\end{fmbox}% comment this for two column layout

%%%%%%%%%%%%%%%%%%%%%%%%%%%%%%%%%%%%%%%%%%%%%%
%%                                          %%
%% The Abstract begins here                 %%
%%                                          %%
%% Please refer to the Instructions for     %%
%% authors on http://www.biomedcentral.com  %%
%% and include the section headings         %%
%% accordingly for your article type.       %%
%%                                          %%
%%%%%%%%%%%%%%%%%%%%%%%%%%%%%%%%%%%%%%%%%%%%%%

\begin{abstractbox}

\begin{abstract} % abstract
\parttitle{Background}
The calculation of diffusion-controlled ligand binding rates is important for understanding enzyme mechanisms as well as designing enzyme inhibitors.
We demonstrate the accuracy and effectiveness of a Lagrangian particle-based method, smoothed particle hydrodynamics (SPH), to study diffusion in biomolecular systems by numerically solving the time-dependent Smoluchowski equation for continuum diffusion.
\parttitle{Results}
The numerical method is first verified in simple systems and then applied to the calculation of ligand binding to an acetylcholinesterase  monomer.
Unlike previous studies, a reactive Robin boundary condition (BC), rather than the absolute absorbing (Dirichlet) boundary condition, is considered on the reactive boundaries.
This new boundary condition treatment allows for the analysis of enzymes with ``imperfect'' reaction rates.
Rates for inhibitor binding to mAChE are calculated at various ionic strengths and compared with experiment and other numerical methods.
We find that imposition of the Robin BC improves agreement between calculated and experimental reaction rates.
\parttitle{Conclusions}
Although this initial application focuses on a single monomer system, our new method provides a framework to explore broader applications of SPH in larger-scale biomolecular complexes by taking advantage of its Lagrangian particle-based nature. 
\end{abstract}

%%%%%%%%%%%%%%%%%%%%%%%%%%%%%%%%%%%%%%%%%%%%%%
%%                                          %%
%% The keywords begin here                  %%
%%                                          %%
%% Put each keyword in separate \kwd{}.     %%
%%                                          %%
%%%%%%%%%%%%%%%%%%%%%%%%%%%%%%%%%%%%%%%%%%%%%%


% MSC classifications codes, if any
%\begin{keyword}[class=AMS]
%\kwd[Primary ]{}
%\kwd{}
%\kwd[; secondary ]{}
%\end{keyword}

\end{abstractbox}
%
%\end{fmbox}% uncomment this for twcolumn layout

\end{frontmatter}

%%%%%%%%%%%%%%%%%%%%%%%%%%%%%%%%%%%%%%%%%%%%%%
%%                                          %%
%% The Main Body begins here                %%
%%                                          %%
%% Please refer to the instructions for     %%
%% authors on:                              %%
%% http://www.biomedcentral.com/info/authors%%
%% and include the section headings         %%
%% accordingly for your article type.       %%
%%                                          %%
%% See the Results and Discussion section   %%
%% for details on how to create sub-sections%%
%%                                          %%
%% use \cite{...} to cite references        %%
%%  \cite{koon} and                         %%
%%  \cite{oreg,khar,zvai,xjon,schn,pond}    %%
%%  \nocite{smith,marg,hunn,advi,koha,mouse}%%
%%                                          %%
%%%%%%%%%%%%%%%%%%%%%%%%%%%%%%%%%%%%%%%%%%%%%%

%%%%%%%%%%%%%%%%%%%%%%%%% start of article main body

\section{Background}

In the ``perfect'' enzyme~\cite{Eigen1974Book} acetylcholinesterase (AChE), the rate-limiting step for catalysis is diffusional encounter \cite{Baze1986,Nolte1980}. 
Specifically, the active site lies at the bottom of a 20 \AA-deep gorge, and the diffusion of substrate into it is accelerated by electrostatic steering \cite{Radic1997,Tan1993}.
Its diffusion-limited behavior, complex geometry, and strong electrostatic influence has made AChE a useful target for both experimental and computational studies of biomolecular diffusion \cite{SongFEM2004,SongMutant2004,Radic1997,Tan1993,Wade1993,Tara1998,Davis1991a,Davis1991b}.

Two major classes of methods have been used to estimate diffusion rates in biomolecular systems.
Mesoscopic coarse-grained methods like Monte Carlo \cite{Genest1989,Saxton1992,Kerr2008}, Brownian Dynamics (BD) \cite{Northrup1988,Wade1993,Tara1998}, and Langevin Dynamics \cite{Eastman1998,Yeomans2001} simulations trace the trajectories of individual coarse-grained particles driven by Brownian motion.
Such simulations typically consider dilute ligand concentrations so that electrostatic protein-ligand interactions can be modeled by the Poisson-Boltzmann equation \cite{Baker2005Book,Baker2005Review} with a few notable exceptions \cite{Senapati2004}.
Alternatively, continuum models can be used to treat the diffusion of ligand concentration in space around a biomolecule by the Smoluchowski equation \cite{Kurnikova1999,Smart1998,Cheng2007,SongFEM2004,LuJCP2010,LuPMC2010}.
In particular, an adaptive finite element approach \cite{Holst:2001a} has been used to numerically solve the Smoluchowski equation, and it shows higher accuracy in predicting experimental data about the ligand binding rates than the coarse-grained BD modeling \cite{SongFEM2004}. 
For dilute ligand concentrations, electrostatic interactions can also be modeled with the Poisson-Boltzmann equation like the mesoscale approach \cite{SongFEM2004,SongMutant2004}.  
However, for more concentrated ligand solutions, continuum models can also model the electrostatic potential near the biomolecular surface using a regularized Poisson-Nernst-Planck formulation \cite{LuJCP2010,LuPMC2010}, allowing screening of the ligand-receptor interactions by its time-dependent distribution around the protein.

Here, we follow the continuum approach but solve the Smoluchowski equation using a new smoothed particle hydrodynamics (SPH) method \cite{Monaghan_SPH_2005,monaghan2012smoothed}.
Unlike Eulerian grid-based methods such as FEM, SPH is a Lagrangian particle-based method.
SPH has been used with good accuracy for numerically solving partial differential equations (PDEs) describing momentum, mass and energy conservation laws \cite{Monaghan_SPH_2005}.
In SPH, the domain is discretized into a set of \lq\lq{}particles\rq\rq{} that serve as interpolation points to numerically solve the governing PDEs.
The SPH discretization of PDEs is based on a meshless interpolation scheme, which allows the PDEs to be written in the form of a system of ordinary differential equations (ODEs).
Unlike grid-based FEM methods, SPH has a straightforward discretization without the need for time-consuming FEM mesh construction around complicated geometries such as biomolecules.
Due to its Lagrangian nature, SPH has many advantages for modeling physical phenomena involving moving boundaries, large deformation of materials, multiphases, and advection-dominated diffusive transport \cite{PanJCP2013,PanIJP2013,monaghan2012smoothed}.
%SPH models advection exactly; there is no need for front tracking schemes because interfaces between phases move with particles; mass, momentum, and energy are explicitly conserved; and complex physics can be included via simple molecular-like interactions.
In addition, the similarity of SPH to molecular dynamics and mesoscopic coarse-grained particle methods (e.g., dissipative particle dynamics, BD, and Langevin dynamics), allows coupling of simulations across scales to build a multiscale modeling framework.
This is our primary goal with the current work:  to enable the multiscale and multiphysics description of biomolecular dynamics and ligand recognition.
To the best of our knowledge, SPH has not been widely used in modeling biomolecular systems.
Thus, in the present work, we aim to take the first step to introduce SPH into this field through the development of a SPH model for biomolecular diffusion with AchE as a test case. 

In the SPH model, the Smoluchowski equation is numerically solved and the ligand binding rates are calculated from flux across the reactive boundary as in  previous studies using FEM \cite{Kurnikova1999,Smart1998,Cheng2007,SongFEM2004,LuJCP2010,LuPMC2010} .
However, in the previous FEM studies, active sites were modeled using the absolute absorbing (Dirichlet) boundary condition (BC).
This BC has a simple description on the reactive boundaries but assumes infinitely fast chemical reactions between the enzyme and the ligand; i.e., a ``perfect enzyme''.
In our model, we take into account imperfect and non-instantaneous reactivity and thus solve the equation using reactive (Robin) boundary condition.

To solve the Smoluchowski equation subject to Robin BC using SPH, we use a continuum surface reaction method \cite{Ryan2010} which we have recently adapted to solve the Navier-Stokes equations subject to slip (Robin) boundary conditions \cite{Pan_CBFSPH_2014}.
In this formulation, the Robin BC is replaced by a reflective Neumann BC and a source term is added into the governing equation.
The derivation of the method is based on the approximation of the sharp boundary with a diffuse interface of finite thickness by means of a color function.
This method is general for any arbitrary complex geometries and thus appropriate for modeling Robin BC in biomolecular systems with complex structures. 

%The accuracy of SPH model for solving the Smoluchowski equation subject to different boundary conditions is first verified on representative spherical test cases. We then apply it to the AchE molecular system and calculate the reaction rates for AchE as a function of time and ionic strength. The results predicted by SPH are compared with experimental data \cite{Radic1997} and other computational methods like BD \cite{Tara1998} and FEM \cite{SongFEM2004}. We find that the present SPH model agrees with the experiment slightly worse than the FEM \cite{SongFEM2004} when the Dirichlet BC is imposed on the reactive boundary. However, the agreement is improved by imposing the Robin BC. Finally, the connections of the results to the biophysics of diffusion-controlled enzymes are discussed.

\section{Results and discussion}
\subsection{Spherical test systems}
\label{sec:verification}

Before the numerical method was applied to a biomolecular system with complicated geometry, we verified it on simple spherical test cases.
Specifically, we considered a diffusing sphere with a radius $R_1$. The entire domain was confined by the outer boundary $\Gamma_b$ determined as a spherical surface with the radius of $R_2 = 125$ {\AA}.
%The Dirichlet BC specified by Eq.~\ref{equ:DirichletBC_outer} was imposed on the outer boundary with the dimensionless bulk concentration ($p_\mathrm{bulk}$) set to 1. 

For the first test case, we let $R_1=0$ and assume no external potential, for which the time-dependent analytical solution of the Smoluchowski equation can be easily derived.
Figure~\ref{fig:sphere_dirichlet_time} compares the SPH numerical solutions with the analytical solution at different times.
SPH solutions are compared at different resolutions and their corresponding $L_2$ errors are calculated relative to the analytical solution.
Figure~\ref{fig:sphere_dirichlet_time} shows that, even at the coarsest resolution ($\Delta x=8$ {\AA}), the SPH solution agrees well with the analytical solution with about 3\% relative error.  This relative error is further reduced to 1\% by increasing the resolution to $\Delta x=2$ {\AA}.       

Next, the spherical system is assumed to have a Coulombic form of the PMF, i.e., $W(r)=q/\beta r$ with +1 $e$ charge. We set $R_1=50$ {\AA} and impose either a Dirichlet BC as specified in Eq.~\ref{equ:DirichletBC_active} or Robin BC as in Eq.~\ref{equ:RobinBC_active}.
In these two tests, the corresponding SPH solutions of concentration at steady-state are compared with the analytical solution. The converged SPH solutions are shown for the Dirichlet BC (Fig.~\ref{fig:sphere_dirichlet_steady}) and Robin BC (Fig.~\ref{fig:sphere_Robin_steady}) imposed on the inner spherical boundary ($r=R_1$). The reactive coefficient for the Robin BC is $\alpha=1\times 10^3$.
In both tests, the SPH solutions show very good agreement with the analytical solution even at the resolution of $\Delta x=8$ {\AA}, which can be further improved with increasing resolution to $\Delta x=2$ {\AA}. 
Moreover, at $\Delta x=2$\AA, the calculated reaction rate is $2.83\times10^{12}\mathrm{M^{-1} {min}^{-1}}$ for the Dirichlet BC, and is $8.24\times10^{11}\mathrm{M^{-1} {min}^{-1}}$ for the Robin BC, both with $L_2$ errors less than 3\% relative to the analytically evaluated ones.

\subsection{Application to acetylcholinesterase-ligand binding rates}
\label{sec:application_AchE}

We applied the SPH method to study the ligand binding kinetics of a simple spherical cationic ligand to acetylcholinesterase (AChE) under various ionic strength conditions.
Specifically, we performed the time-dependent calculations at ionic strengths of 0.0, 0.05, 0.10, 0.15, 0.20, 0.50 and 0.67 M until the diffusion reaches the steady-state.
To achieve the highest accuracy with affordable computational cost, a resolution of $\Delta x=2$ {\AA} was used in all the following calculations. 

In previous studies by Song et al. \cite{SongFEM2004}, a simple but realistic set of boundaries was used inspired by Tara et al.\  \cite{Tara1998}, encompassing the active site as well as the gorge and the peripheral anionic site (PAS) of AChE.
We constructed these spherical active boundaries ($\Gamma_a$) at varying distances from the active site along an axis defined by the carbonyl carbon of S203 at the origin and the gorge.
Spheres 1-6 were placed at 16.6, 13.6, 10.6, 7.6, 4.6, and 1.6 {\AA} along the this axis, respectively.
The outermost spheres 1 and 2 were assigned radii of 12 and 9 {\AA}, respectively, while all others were given radii of 6 {\AA}. Figure~\ref{fig:4B82mono_domain}A shows the discretized domain with $R_2 = 128$ {\AA}. Figure~\ref{fig:4B82mono_domain}B and \ref{fig:4B82mono_domain}C depict the constructed reactive boundaries 1 and 4.

In most prior studies \cite{SongFEM2004,SongMutant2004,Cheng2007}, an absolute absorbing (Dirichlet) boundary condition (Eq.~\ref{equ:DirichletBC_active}) was assumed.
However, in the present work, we demonstrated improved performance with the reactive (Robin) boundary condition (Eq.~\ref{equ:RobinBC_active}) imposed on the reactive boundaries. 
Figure~\ref{fig:4B82mono_concen_ionic} shows the steady-state spatial distribution of ligand throughout the simulation domain at different ionic strengths.
At zero ionic strength, there are three large ligand-attracting regions, two on either side of the active site and one on the opposite side of the protein.
There is also one ligand-depleted region at the top and another one near the opening of the gorge.
At non-zero ionic strengths, electrostatic screening reduces the size of the ligand-enriched and ligand-depleted regions.
However, a large region around the active site remains ligand-depleted at up to 0.50 M ionic strength. 
Figure~\ref{fig:4B82mono_concen_015M_time} illustrates the temporal evolution of the concentration distribution as ligand moves inward in the bulk region from the outer boundary ($\Gamma_b$). The distribution has clearly reached steady state by 190 ns. 

We calculated the reaction rates from these solutions according to Eq.~\ref{equ:reaction_rate_SPH_Robin}. In Figure~\ref{fig:4B82mono_k_ionic}, the left panel shows the time evolution of $k_\mathrm{on}(t)$ on reactive boundary 1 at different ionic strengths.
For this boundary, $k_\mathrm{on}(t)$ converges within 150 ns for all ionic strengths. 
The right panel shows $k_\mathrm{on}(t)$ on reactive boundaries 1-4, respectively, at 0.15 M ionic strength.

We have quantitatively compared the reaction rates calculated by SPH with experiment \cite{Radic1997} and previous computational studies by FEM \cite{Wade1993,SongFEM2004}.
 Radic et al. \cite{Radic1997} fit their experimentally measured reaction rates as a function of ionic strength using the Debye-H\"uckel limiting law
\begin{equation}
	k_\mathrm{on}= (k_\mathrm{on}^0-k_\mathrm{on}^\mathrm{H})10^{-1.18|Z_E Z_I|\sqrt{I}}+k_\mathrm{on}^\mathrm{H},
	\label{equ:Debye_Huckel}
\end{equation}
where $I$ is the ionic strength, $k_\mathrm{on}^0$ is the effective reaction rate at zero ionic strength rate, $k_\mathrm{on}^\mathrm{H}$ is the effective limiting reaction rate at infinite ionic strength and set to the value of $k_\mathrm{on}$ calculated at 0.67 M ionic strength, $z_E$ is the effective enzyme charge, and $z_I$ is the effective inhibitor charge with a fixed value of +1 $e$. For the Robin BC SPH calculations, the reaction coefficient $\alpha$ was varied, as shown in Figure~\ref{fig:4B82mono_k_alpha}, to identify the value of $8.0 \times 10^3$ which optimized agreement between computational and experimental results.

Figure~\ref{fig:4B82mono_k_datacomp} and  Table~\ref{tab:4B82mono_k_datacomp} compare the reaction rates from SPH, FEM \cite{Wade1993,SongFEM2004}, BD, and experimental data by Radic et al. \cite{Radic1997}.
%The reaction rates calculated by Eq.~\ref{equ:reaction_rate_SPH_Dirichlet} given the Dirichlet BC is imposed on reactive boundaries are also presented in Figure~\ref{fig:4B82mono_k_datacomp}A to elucidate the effect of boundary conditions on reaction rates.
As noted by Song et al. (2004) \cite{SongFEM2004}, BD simulations systematically overestimate the experimental $k_\mathrm{on}$, while the FEM produces good agreement with experimental $k_\mathrm{on}$ at RMSD = 0.37 $\mathrm{M^{-1} {min}^{-1}}$.
With the Dirichlet BC, SPH predicts $k_\mathrm{on}$ with RMSD of 0.57 $\mathrm{M^{-1} {min}^{-1}}$, intermediate between FEM and BD results.
However, with the Robin BC, SPH predicts $k_{\text{on}}$ with RMSD of 0.33 $\mathrm{M^{-1} {min}^{-1}}$, better than the FEM and BD results.

We also assessed the accuracy of SPH method for describing the ligand-binding kinetics of a mAchE surface mutant. We tested the surface hexa-mutant (E84Q, E91Q, D280V, D283N, E292Q, and D372N) from Radic et al.\ \cite{Radic1997}, which reduces the reaction rate by about a factor of 4 across the 0 to 0.67 M ionic strengths.  For the mutants, which are nearly isosteric with the wild-type protein, we used the same SPH model as the wild type, but recalculated the electrostatic potentials for the mutant charge distribution. As presented in Table~\ref{tab:4B82mono_k_datacomp}, the Robin BC SPH model has qualitative accuracy:  predicting $k_\mathrm{on}^0$ of 2.01 compared to 1.80 from Radic et al. \cite{Radic1997} with a $k_\mathrm{on}$ RMSD of 0.70 over the entire ionic strength range studied.

\section{Conclusions}
\label{sec:discussion}
The Robin BC offers a new way to incorporate reactive surfaces into continuum diffusion models for rate calculations.
This Robin-based model incorporates a new parameter $\alpha$, which has units of \AA/$\rm{\mu}$s and can be related to the probability of reaction within distance $\Delta x$ of the boundary and time interval $\Delta t$ by $P=1-\exp(-\alpha\frac{\Delta t}{\Delta x})$ \cite{Gillespie2007}.
Thus, $\alpha=0$ corresponds to zero reactivity (reflective Neumann boundary conditions) while $\alpha=\infty$ corresponds to absolute reactivity (absorbing Dirichlet boundary conditions).

There are two possible origins for the differences between the current SPH model results and past FEM calculations using the Dirichlet BC.  
First, the current SPH work uses a more recent AChE structure (4B82) while the previous FEM calculations used an older structure (1MAH).
Second, our SPH model uses a fixed resolution uniformly on both solution domain and boundaries, while the FEM adaptively meshes the reactive boundary with higher resolution.

This work has provided an initial demonstration that the Lagrangian (particle-based) SPH method out-performs the Eulerian (grid-based) finite element method \cite{SongFEM2004} in accurately predicting ligand binding rates in AChE.
This result is important because while both methods can be used to study molecules of the size of AChE, SPH is more scalable to larger systems such as the synapse geometry where AChE operates.
Additionally, due to its Lagrangian nature, SPH can easily incorporate other physical phenomena such as fluid flow or protein flexibility.
%Thus, it can account for allosteric coupling between ligand approach to the binding site and the reactivity of the protein conformation \cite{Cai2011,Zhou2010}. Moreover, hydrodynamics in the environment of biomolecules can be easily incorporated in SPH to investigate the advection-driven diffusive transport in biomolecular systems.

We have demonstrated that superior performance can be achieved using a probabilistic reactive (Robin) boundary condition rather than a simple Dirichlet BC.
In fact, the Robin BC is likely more biologically relevant than the Dirichlet BC.
While the AChE enzyme is considered nearly \lq\lq{}perfect\rq\rq{} with a diffusion-limited reaction rate, there is experimental evidence that a very small fraction of substrates entering the active site gorge do not react. Specifically, recent kinetic experiments suggest that through unknown mechanisms, the PAS limits the rate of progression of non-substrates of any size to the catalytic site \cite{Beri2013}. In addition, molecular dynamics simulations suggest that the peripheral anionic site (PAS) provides a selective gating function, for example by fluctuations in the gorge width that are likely to let acetylcholine but not let larger molecule pass through \cite{Zhou1998,Baker1999}.

\section{Methods}

\subsection{Governing equations and boundary condition}
\label{subsec:governing_equs_boundary condition}

The time-dependent Smoluchowski equation can be written as:
\begin{equation}
	\frac{d p(\mathbf{x},t)}{dt}=\nabla\cdot \mathbf{J} (\mathbf{x},t), ~~~~\mathbf{x}\in\Omega,
	\label{equ:Smol}
\end{equation}
where $p(\mathbf{x},t)$ is the concentration distribution of the reactants, and the concentration flux $ \mathbf{J} (\mathbf{x},t)$ is defined as:
\begin{equation} 
	\mathbf{J} (\mathbf{x},t) = D (\mathbf{x})[\nabla p(\mathbf{x},t) + \beta p(\mathbf{x},t) \nabla W(\mathbf{x})],
	\label{equ:flux}
\end{equation}
where $D (\mathbf{x})$ is the diffusion coefficient; for simplicity, this is assumed to be constant.
$\beta=1/k_B T$ is the inverse Boltzmann energy with the Boltzmann constant $k_B$ and kinetic temperature $T$.
$W(\mathbf{x})$ is the potential mean force (PMF) for the diffusing particle due to solvent-mediated interactions with the target molecule.  
The equation is solved in a three-dimensional domain $\Omega$, subject to the following boundary conditions.
First,
\begin{equation}
	p(\mathbf{x},t)= p_{\text{bulk}} ~\text{for} ~\mathbf{x}\in\Gamma_b,
	\label{equ:DirichletBC_outer}
\end{equation}
specifying a Dirichlet BC on the outer boundary $\Gamma_b$ where the concentration is equal to a bulk concentration $p_{\text{bulk}}$.
The outer boundary is often a spherical surface with a radius chosen to ensure that the ligand-protein potential is spherically symmetric and/or can be approximated analytically \cite{SongFEM2004}.
For the current study with AChE, this outer boundary has radius $R_2 \approx 128 \mathrm{\AA}$ as determined following a procedure similar to Song et al and Chen et al \cite{SongFEM2004,Cheng2007}.
%To test the adequacy of the simulation domain size, we compared the electrostatic potential (see Section \ref{subsec:calculation_pmf}) on the edge of a $257 \mathrm{\AA} \times 257 \mathrm{\AA} \times 257 \mathrm{\AA}$ box, which is enough to enclose the entire simulation domain, with the potential at the same position calculated from a  $400 \mathrm{\AA} \times 400 \mathrm{\AA} \times 400 \mathrm{\AA}$ box.
%For all ionic strengths, the potential on the edge of the smaller box differs by at most 0.001 $k_B T/e$ from the equivalent position in the larger box.
%The difference ranges from 0.001 at 0.0 ionic strength to 1e-12 at 0.50 M.
%This indicates that the potential has converged to bulk values by the outer boundary of the simulation domain with $R_2 \approx 128 \mathrm{\AA}$.
Also following Song et al and Chen et al, $p$ is normalized such that $p_{\mathrm{bulk}} = 1$.

The active site boundary $\Gamma_a$ was modeling using either reactive Robin or absolute absorbing Dirichlet BCs:
\begin{equation}
	\mathbf{n}(\mathbf{x}) \cdot \mathbf{J} (\mathbf{x},t) = \alpha p(\mathbf{x},t) ~ \text{for} ~ \mathbf{x}\in\Gamma_a,
	\label{equ:RobinBC_active}
\end{equation}
or
\begin{equation}
	p(\mathbf{x})= 0 ~ \text{for} ~ \mathbf{x}\in\Gamma_a,
	\label{equ:DirichletBC_active}
\end{equation}
respectively.
The coefficient $\alpha$ is chosen to model an intrinsic reaction rate for the active site.
Finally, a reflective Neumann BC is defined on the non-reactive boundary of molecule
\begin{equation}
	\mathbf{n}(\mathbf{x}) \cdot \mathbf{J} (\mathbf{x},t) = 0 ~ \text{for} ~ \mathbf{x}\in\Gamma_m.
	\label{equ:NeumannBC_molecule}
\end{equation}
Figure~\ref{fig:solution_domain} shows the  simulation domain along with all boundaries.

Given a solution to Eq.\ \ref{equ:Smol}, the reaction rate is calculated from the integral of the flux across the reactive surface \cite{Zhou1990}:
\begin{equation}
	k_{\text{on}}= p_\mathrm{bulk}^{-1} \iint\limits_{\Gamma_a} \mathbf{n}(\mathbf{x}\rq{}_s) \cdot \mathbf{J} (\mathbf{x}\rq{}_s,t) d\mathbf{x}'_s.
	\label{equ:reaction_rate}
\end{equation}

In order to solve the Smoluchowski equation (Eq.~\ref{equ:Smol}) subject to the reactive Robin BC (Eq.~\ref{equ:RobinBC_active}), the simulation domain is extended to include a sub-domain $\Omega_a$ that is separated from $\Omega$ by $\Gamma_a$, and we then reformulate Eq.~\ref{equ:Smol} as:
\begin{align}
	\frac{dp^r(\mathbf{x},t)}{dt}= &\nabla\cdot \left(D (\mathbf{x})[\nabla p^r(\mathbf{x},t) + \beta p^r(\mathbf{x},t) \nabla W(\mathbf{x})]\right) \nonumber \\
	- &\alpha p^r (\mathbf{x}, t) \iiint\limits_{\Omega_a} [\mathbf{n}(\mathbf{x}) + \mathbf{n}(\mathbf{x}')]\cdot \nabla_{\mathbf{x}} w(\mathbf{x}-\mathbf{x}', h_r) d\mathbf{x}', ~~~\mathbf{x}\in\Omega,
	\label{equ:smol_csf_final}
\end{align}
subject to the reflective Neumann BC:
\begin{equation}
	\mathbf{n}(\mathbf{x}) \cdot \mathbf{J}^r (\mathbf{x},t) = 0 ~ \text{for} ~ \mathbf{x}\in\Gamma_a.
	\label{equ:NeumannBC_active}
\end{equation}
The derivation of Eq.~\ref{equ:smol_csf_final} is detailed in Appendix~\ref{sec:RobinBC_formulation_derivation}, which demonstrates
\begin{equation}
	\lim_{h_r \to 0} p^r(\mathbf{x},t) = p(\mathbf{x},t).
\end{equation}
In Eq.~\ref{equ:smol_csf_final}, the normalized kernel function, $w(\mathbf{x})$, is a positive bell-shaped function with at least first continuous derivative and compact support $\kappa h_r$, $w(|\mathbf{r}|>\kappa h_r) = 0$.
The value of $\kappa$ depends on the specific functional form of $w(\mathbf{x})$, which is specified in Section \ref{subsec:SPH_formulation}.
In particular, $w(\mathbf{x})$ satisfies the following conditions:
\begin{equation}
	\iiint\limits_{\Omega\cup\Omega_a}  w(\mathbf{x}-\mathbf{x}', h_r) d\mathbf{x}' =1
	\label{equ:kernel_func_integ}
\end{equation}
and
\begin{equation}
	\lim_{h_r\to 0}w(\mathbf{x}-\mathbf{x}', h_r)=\delta(\mathbf{x}-\mathbf{x}').
	\label{equ:kernel_delta_func}
\end{equation}
The normal unit vector $\mathbf{n}$ in Eq.~\ref{equ:smol_csf_final} can be found in terms of a smoothed color function $\tilde{\phi}$ as defined in Appendix~\ref{sec:RobinBC_formulation_derivation}:
\begin{equation}
	\mathbf{n}(\mathbf{x}) = \frac{\nabla \tilde{\phi}(\mathbf{x})}{|\nabla \tilde{\phi}(\mathbf{x})|},  ~~~ \mathbf{x} \in \Omega\cup\Omega_a.	
	\label{equ:normal_vector}
\end{equation}

\subsection{SPH discretization of equations and boundary conditions}
\label{subsec:SPH_formulation}

In this section, we present SPH discretization of the Smoluchowski equation, using Eq.~\ref{equ:Smol} if the Dirichlet BC is used and Eq.~\ref{equ:smol_csf_final} if the Robin BC is assumed.
To simplify notation, we omit superscript $r$ for the variables in Eq.~\ref{equ:smol_csf_final} in the subsequent derivations.

The domain $\Omega$, and the boundaries $\Gamma_a$ and $\Gamma_b$ (extended as domains $\Omega_a$ and $\Omega_b$ respectively), are discretized with a set of N points with positions denoted by a vector $\mathbf{r}_i$ ($i=1,...,N$).
The points (which are commonly referred to as particles in SPH) are used to discretize and solve the governing equation.
Initially,  the particles are distributed uniformly (e.g., placed on a regular cubic lattice) with $d_i$ as the prescribed number density at $\mathbf{r}_i$.
The discretization is based on a meshless interpolation scheme:
\begin{equation}
	A_i \approx \sum\limits_{j} \frac{A_j}{d _j}w(\mathbf{r}_{ij},h),
	\label{equ:SPH_A_Interp}
\end{equation}
where $A_i = A(\mathbf{r}_i)$ is a function defined at particle $i$, $\mathbf{r}_{ij}=\mathbf{r}_i-\mathbf{r}_j$ and $w(\mathbf{r}_{ij},h)$ is the weighting kernel function. The interpolation scheme assumes a summation over all neighboring SPH particles but, due to the compact support of $w$, only particles within distance $\kappa h$ from $\mathbf{r}_i$ have a non-zero contribution to the summation.
Spatial derivatives of  $A$ can be calculated as
\begin{equation}
	\nabla_i A_i \approx \sum\limits_{j} \frac{A_j}{d _j} \nabla_i w( \mathbf{r}_{ij},h ).
	\label{equ:SPH_A_Interp_D}
\end{equation}

In the present work, we  use a cubic spline kernel as the weighting function
\begin{equation}
	w(\mathbf{r},h) = \frac{1}{\pi h^3}
	\begin{cases}
		1-\frac{3}{2}q^2+\frac{3}{4}q^3 & ~~~0 \leq q \leq 1 \\ 
		\frac{1}{4}(2-q)^3 & ~~~1 < q \leq 2 \\
		0 & ~ q>2,
	\end{cases} 
	\label{equ:cubic_kernel}
\end{equation}
where $q=|{\bf r}|/h$.
With this form of weighting function, only particles within $2h$ distance from particle $i$ contribute to the summations in the SPH equations.
We have chosen $h=1.3\Delta x$ where $\Delta x$ is the size of the cubic lattice.   

The SPH approximation of functions and their spatial derivatives allows the Smoluchowski equation subject to the Dirichlet BC (Eq.~\ref{equ:Smol}) to be written as a ODE governing the evolution of concentration on particle $i$ as:
\begin{multline}
	\frac{d p_i}{dt} =  \sum_{j\in \Omega\cup\Omega_b\cup\Omega_a} \frac{D_i+D_j}{d_j} (p_i-p_j) \frac{1}{r_{ij}} \frac{d w(r_{ij}, h)}{d r_{ij}} \\
	+  \beta \sum_{j\in \Omega} \frac{D_i p_i+D_j p_j}{d_j} (W_i-W_j) \frac{1}{r_{ij}} \frac{d w(r_{ij}, h)}{d r_{ij}} .
	\label{equ:SPH_discretize_Dirichlet}
\end{multline}
The derivations of the first and second terms on the right-hand side of Eq.~\ref{equ:SPH_discretize_Dirichlet} can be found in Monaghan et al.\ \cite{Monaghan_SPH_2005} where $r_{ij}$ is the magnitude of the vector $ \mathbf{r}_{ij}$.
If the reactive Robin BC on reactive boundary is imposed, Eq.~\ref{equ:smol_csf_final} is then solved instead and its corresponding SPH discretization form is: 
\begin{multline}
	\frac{d p_i}{dt} =  \sum_{j\in \Omega\cup\Omega_b} \frac{D_i+D_j}{d_j} (p_i-p_j) \frac{1}{r_{ij}} \frac{d w(r_{ij}, h)}{d r_{ij}} \nonumber\\
	+ \beta \sum_{j\in \Omega} \frac{D_i p_i+D_j p_j}{d_j} (W_i-W_j) \frac{1}{r_{ij}} \frac{d w(r_{ij}, h)}{d r_{ij}} \\
	-  \alpha p_i \sum_{k\in \Omega_a} \frac{\mathbf{n}_i + \mathbf{n}_k}{d_k}\cdot \frac{\mathbf{r}_{ik}}{r_{ik}} \frac{d w(r_{ik}, h)}{d r_{ik}}.
	\label{equ:SPH_discretize_Robin}
\end{multline}
The last term in Eq.~\ref{equ:SPH_discretize_Robin} is obtained by discretizing the integral in Eq.~\ref{equ:smol_csf_final} as a Riemann sum:
\begin{multline}
	\alpha p(\mathbf{x}, t)  \iiint\limits_{\Omega_a} [\mathbf{n}(\mathbf{x}) + \mathbf{n}(\mathbf{x}')]\cdot \nabla_{\mathbf{x}} w(\mathbf{x}-\mathbf{x}', h_r) d\mathbf{x}' \\
	= \alpha p(\mathbf{x}, t) \sum_{k \in \Omega_a}
  \Delta V_k [\mathbf{n}(\mathbf{x}) + \mathbf{n}_k] \cdot \nabla_{\mathbf{x}} w(\mathbf{x} - \mathbf{r}_{k}, h_r), 
\end{multline}
where $\Delta V_k = \frac{1}{d_k}$ is the volume of particle $k$ and $\sum_{k \in \Omega_a}$ is the summation over the reactive boundary particles within $2h_r$ distance from particle $i$. 
The SPH expression for calculating the normal unit vector is obtained as:
\begin{equation}
	\mathbf{n}_{i} = \frac{\sum\limits_{j \in \Omega\cup\Omega_a}\frac{1}{d_j}(\phi_j-\phi_i)\nabla_{i} w(\mathbf{r}_{ij}, h_r)}{\left |\sum\limits_{j \in \Omega\cup\Omega_a}\frac{1}{d_j}(\phi_j-\phi_i)\nabla_{i} w(\mathbf{r}_{ij}, h_r) \right |}.
\label{equ:SPH_normal_vector}
\end{equation}
In the simulations presented below, we set $h_r = h$ but it could be set differently in future applications.

Note that the reflective Neumann BC (Eq. \ref{equ:NeumannBC_molecule} or \ref{equ:NeumannBC_active}) can be simply enforced in SPH by excluding the contribution from the boundary particles in the summation.
The Dirichlet BC (Eq. \ref{equ:DirichletBC_outer} or \ref{equ:DirichletBC_active}) is enforced by assigning the fixed boundary value of concentration on the boundary particles.   
If the Robin BC is imposed, the reaction rate $k_{\mathrm{on}}(t)$ can be calculated by Eq.~\ref{equ:reaction_rate_final_Robin} in Appendix~\ref{sec:calc_reaction_rate} and its corresponding SPH discretization form is:  
\begin{equation}
	k_{\mathrm{on}}= \sum_{i \in \Omega} \alpha p_i \left[\sum_{k\in \Gamma_a} \frac{\mathbf{n}_i + \mathbf{n}_k}{d_k}\cdot \frac{\mathbf{r}_{ik}}{r_{ik}} \frac{d w(r_{ik}, h)}{d r_{ik}}\right].
	\label{equ:reaction_rate_SPH_Robin}
\end{equation}
Otherwise, when the Dirichlet BC is enforced on the reactive boundary, the discretization of Eq.~\ref{equ:reaction_rate_final} in Appendix~\ref{sec:calc_reaction_rate} is:  
\begin{equation}
	k_{\mathrm{on}}= \sum_{i \in \Omega} (\mathbf{n}_i \cdot \mathbf{J}_i) \left[\sum_{k\in \Gamma_a} \frac{\mathbf{n}_i + \mathbf{n}_k}{d_k}\cdot \frac{\mathbf{r}_{ik}}{r_{ik}} \frac{d w(r_{ik}, h)}{d r_{ik}}\right],
	\label{equ:reaction_rate_SPH_Dirichlet}
\end{equation}
where 
\begin{equation}
	\mathbf{J}_i = D_i \sum_{j\in \Omega\cup\Omega_b\cup\Omega_a} \frac{p_j-p_i}{d_j}\frac{\mathbf{r}_{ij}}{r_{ij}} \frac{d w(r_{ij}, h)}{d r_{ij}} + \beta D_i p_i \sum_{j\in \Omega} \frac{W_j-W_i}{d_j} \frac{\mathbf{r}_{ij}}{r_{ij}} \frac{d w(r_{ij}, h)}{d r_{ij}}.
	\label{equ:fluxJ_SPH}
\end{equation}

\subsection{Calculation of potentials of mean force}
\label{subsec:calculation_pmf}

We calculated the potential of mean force $W(\mathbf{x})$ using the recently published 2.1 {\AA} resolution structure of mouse AChE \cite{Andersson2013}.
To prepare this structure for the calculation, we assigned titration states of ionizable residues using PROPKA \cite{Li2005} at pH 7, and we used PDB2PQR v1.8 \cite{Dolinsky2007,Dolinsky2004} to assign atomic radii and charges. APBS v1.4 was used to perform a nonlinear Poisson-Boltzmann multi-grid calculation of the electrostatic potential over the entire simulation domain \cite{BakerPNAS2001}.
The small and large domains were set to 600 {\AA} and 400 {\AA}, respectively, with a fine grid spacing of 0.600 {\AA}.
For APBS calculations, we used the single Debye-H\"uckel boundary condition, a smoothed molecular surface, and protein and solvent dielectrics of 2 and 78.54, respectively.
Atomic charges were mapped onto the grids using cubic B-spline discretization.
The calculated potential was mapped onto the SPH discretization points of protein and solvent via trilinear interpolation.

%%%%%%%%%%%%%%%%%%%%%%%%%%%%%%%%%%%%%%%%%%%%%%
%%                                          %%
%% Backmatter begins here                   %%
%%                                          %%
%%%%%%%%%%%%%%%%%%%%%%%%%%%%%%%%%%%%%%%%%%%%%%

\begin{backmatter}

\section*{Competing interests}
  The authors declare that they have no competing interests.

\section*{Author contributions}
WP implemented the new SPH methods, performed simulations, and participated in the writing of the manuscript.
MD performed simulations, and participated in the writing of the manuscript.
NAB conceived of the study, participated in its design and coordination, and helped to draft the manuscript.
All authors read and approved the final manuscript.

\section*{Acknowledgments}
The authors gratefully acknowledge the funding support by the Applied Mathematics Program within the U.S.~Department of Energy Office of Advanced Scientific Computing Research as part of the Collaboration on Mathematics for Mesoscopic Modeling of Materials (CM4). The research was performed using PNNL Institutional Computing at Pacific Northwest National Laboratory. The Pacific Northwest National Laboratory is operated for the U.S. Department of Energy by Battelle under Contract DE-AC06-76RL01830.
This research was also supported by National Institutes of Health grant GM069702 to NAB.

\begin{appendices}

\section{Continuum surface reaction method}
\label{sec:RobinBC_formulation_derivation}
% nab stopped here

In this appendix, we present a detailed derivation of the continuum surface reaction method for solving the Smoluchowski equation subject to Robin BC.
We start from a two-sided problem; i.e., the concentration field $p(\mathbf{x},t)$ is extended into the sub-domain $\Omega_a$ that is separated from $\Omega$ by $\Gamma_a$ such that Eq.~\ref{equ:Smol} can be approximated as
\begin{equation}
	\frac{dp^r(\mathbf{x},t)}{dt}= \nabla\cdot \left(D (\mathbf{x})[\nabla p^r(\mathbf{x},t) + \beta p^r(\mathbf{x},t) \nabla W(\mathbf{x})]\right)  - P_{\Omega}(\mathbf{x},t) ~ \text{for} ~ \mathbf{x}\in\Omega\cup\Omega_a
	\label{equ:smol_csf_reform1}
\end{equation}
subject to
\begin{equation}
	\mathbf{n}(\mathbf{x}_s) \cdot [\mathbf{J}^r(\mathbf{x}_s,t)|_{\mathbf{x}_s\in\Gamma^F_a}-\mathbf{J}^r(\mathbf{x}_s,t)|_{\mathbf{x}_s\in\Gamma^S_a}]= 0 ~ \text{for} ~  \mathbf{x}_s\in\Gamma_a,
	\label{equ:NeumannBC_twosided}
\end{equation}
where $\Gamma^F_a$ and $\Gamma^S_a$ are the two sides of $\Gamma_a$, respectively.
The boundary condition Eq.~\ref{equ:NeumannBC_twosided} emphasizes that the extended concentration field is continuous across $\Gamma_a$.
Comparison of the weak formulations of Eq.~\ref{equ:Smol} subject to Eq.~\ref{equ:RobinBC_active} and Eq.~\ref{equ:smol_csf_reform1} subject to Eq.~\ref{equ:NeumannBC_twosided} yields the relationship
\begin{equation}
	\iiint\limits_{\Omega\cup\Omega_a} P_{\Omega} (\mathbf{x},t) d\mathbf{x} = \iint\limits_{\Gamma_a} \alpha p(\mathbf{x}'_s,t) d \mathbf{x}'_s.
\end{equation}
This weak formulation is obtained by integrating the momentum equations over their respected domains and then applying Gauss' theorem with the corresponding boundary conditions.
To derive the formulation of $P_{\Omega}$, we define a color function (i.e., a sharp characteristic function) as: 
\begin{equation}
	\phi(\mathbf{x}) =
		\begin{cases}
	     0, & \mathbf{x}\in\Omega,\\
    	 1, & \mathbf{x}\in\Omega_a.
   	\end{cases}
	\label{equ:color_func}
\end{equation}
and its smooth counterpart as
\begin{equation}
	\tilde{\phi}(\mathbf{x}) = \iiint\limits_{\Omega\cup\Omega_a} \phi(\mathbf{x}') W(\mathbf{x}-\mathbf{x}', h_r) d\mathbf{x}'.
	\label{equ:smooth_phi}
\end{equation}
The gradient of $\tilde{\phi}$ can then be found from Eq.~\ref{equ:smooth_phi} as
\begin{equation}
	\nabla \tilde{\phi}(\mathbf{x}) = \iiint\limits_{\Omega\cup\Omega_a} \phi(\mathbf{x}') \nabla_{\mathbf{x}} W(\mathbf{x}-\mathbf{x}', h_r) d\mathbf{x}'.
	\label{equ:smooth_phi_gradient}
\end{equation}
Using the definition of the surface delta function \cite{lange2012potential}:
\begin{equation}
	\delta [\mathbf{n}(\mathbf{x}_s) \cdot (\mathbf{x}_s-\mathbf{x})] = \mathbf{n}(\mathbf{x}) \cdot \nabla {\phi}(\mathbf{x}),  ~~~ \mathbf{x} \in \Omega\cup\Omega_a,  ~\text{for}~  \mathbf{x}_s\in\Gamma_a,
	\label{equ:surface_delta}
\end{equation}
and noting that 
\begin{equation}
	\lim_{h_r\to 0} \tilde{\phi} = \phi,
\end{equation}
we can rewrite the surface delta function in terms of $\tilde{\phi} $ as:
\begin{equation}
	\delta [\mathbf{n}(\mathbf{x}_s) \cdot (\mathbf{x}_s-\mathbf{x})] = \mathbf{n}(\mathbf{x}) \cdot \lim_{h_r\to 0} \nabla \tilde{\phi}(\mathbf{x}),  ~\text{for}~ \mathbf{x} \in \Omega\cup\Omega_a, ~  \mathbf{x}_s\in\Gamma_a.
	\label{equ:surface_delta2}
\end{equation}
The surface integral can then be rewritten as a volume integral through the surface delta function:
\begin{align}
	\iint\limits_{\Gamma_a} \alpha p(\mathbf{x}'_s, t) d\mathbf{x}'_s
	&=  \iiint\limits_{\Omega\cup\Omega_a} \alpha p(\mathbf{x},t) \delta [\mathbf{n}(\mathbf{x}_s) \cdot (\mathbf{x}_s-\mathbf{x})] d\mathbf{x},  ~\text{for}~  \mathbf{x}_s\in\Gamma_a, \nonumber \\
	& =  \iiint\limits_{\Omega\cup\Omega_a} \alpha p^r(\mathbf{x},t) \mathbf{n}(\mathbf{x}) \cdot  \nabla \tilde{\phi}(\mathbf{x})  d\mathbf{x}.
	\label{equ:f_surface}
\end{align}
To uniquely define $P_{\Omega}(\mathbf{x},t)$, we require it to vanish at a normal distance greater than $h_r$ from $\Gamma_a$ and require that 
\begin{equation}
	\lim_{h_r\to 0}\iiint\limits_{\Omega\cup\Omega_a} P_{\Omega} (\mathbf{x},t) d\mathbf{x}
	=\iint\limits_{\Gamma_a} \alpha p(\mathbf{x}'_s,t) d\mathbf{x}'_s .
	\label{equ:f_volume_Gauss}
\end{equation}
Comparing Eqs.~\ref{equ:f_surface} and \ref{equ:f_volume_Gauss} yields an expression for $P_{\Omega}(\mathbf{x},t)$ as:

\begin{equation}
	P_{\Omega}(\mathbf{x},t)  = \alpha p^r(\mathbf{x},t) \mathbf{n}(\mathbf{x}) \cdot  \nabla \tilde{\phi}(\mathbf{x}) , ~\text{for}~ \mathbf{x} \in \Omega\cup\Omega_a.
	\label{equ:f_volume}
\end{equation}
Eq.~\ref{equ:smol_csf_reform1} can then be rewritten by combining Eqs.~\ref{equ:f_volume} and \ref{equ:smooth_phi_gradient} as:
\begin{multline}
	\frac{dp^r(\mathbf{x},t)}{dt} = \nabla\cdot \left(D (\mathbf{x})[\nabla p^r(\mathbf{x},t) + \beta p^r(\mathbf{x},t) \nabla W(\mathbf{x})]\right) \\
	- \alpha p^r (\mathbf{x}, t) \iiint\limits_{\Omega\cup\Omega_a} \mathbf{n}(\mathbf{x}) \cdot [\phi(\mathbf{x}') \nabla_{\mathbf{x}} w(\mathbf{x}-\mathbf{x}', h_r)] d\mathbf{x}', 
	~\text{for}~\mathbf{x}\in\Omega\cup\Omega_a.
	\label{equ:smol_csf_reform2}
\end{multline}
Since $p^r$ is not uniquely defined on $\Omega_a$, we introduce a one-sided formulation by approximating Eq.~\ref{equ:smol_csf_reform2} as:
\begin{multline}
	\frac{dp^r(\mathbf{x},t)}{dt} = \nabla\cdot \left(D (\mathbf{x})[\nabla p^r(\mathbf{x},t) + \beta p^r(\mathbf{x},t) \nabla W(\mathbf{x})]\right) \\
	- \alpha p^r (\mathbf{x}, t) \iiint\limits_{\Omega\cup\Omega_a} [\mathbf{n}(\mathbf{x})+\mathbf{n}(\mathbf{x}\rq{})] \cdot [\phi(\mathbf{x}') \nabla_{\mathbf{x}} w(\mathbf{x}-\mathbf{x}', h_r)] d\mathbf{x}', ~\text{for}~\mathbf{x}\in\Omega,
	\label{equ:smol_csf_oneside}
\end{multline}
subject to the reflective Neumann boundary condition (Eq.~\ref{equ:NeumannBC_active}).
Note that $\phi$ is non-zero only in $\Omega_a$, where it is equal to 1 as defined in Eq.~\ref{equ:color_func}.
Thus, the modified governing equation takes its final form as Eq.~\ref{equ:smol_csf_final}.

\section{Calculation of Reaction Rate}
\label{sec:calc_reaction_rate}

Similar to the derivation in Eq.~\ref{equ:f_surface}, using the definition of the surface delta function and given $p_\mathrm{bulk}=1$, the expression for the reaction rate can be rewritten as
\begin{equation}
	k_{\text{on}}=\iiint\limits_{\Omega\cup\Omega_a} [\mathbf{n}(\mathbf{x}) \cdot \mathbf{J} (\mathbf{x},t)] [\mathbf{n}(\mathbf{x})  \cdot  \nabla \tilde{\phi}(\mathbf{x})] d\mathbf{x}.  
	\label{equ:reaction_rate_reform1}
\end{equation}
Substituting Eq.~ \ref{equ:smooth_phi_gradient} into the above equation and using Eq.~\ref{equ:color_func}, a new expression of $k_{\text{on}}$ can be obtained:
\begin{equation}
	k_{\text{on}}=\iiint\limits_{\Omega\cup\Omega_a} [\mathbf{n}(\mathbf{x}) \cdot \mathbf{J} (\mathbf{x},t)]  \iiint\limits_{\Omega_a}  \mathbf{n}(\mathbf{x})  \cdot \nabla_{\mathbf{x}} w(\mathbf{x}-\mathbf{x}', h_r) d\mathbf{x}' d\mathbf{x}.  
	\label{equ:reaction_rate_reform2}
\end{equation}
Similar to Eq.~\ref{equ:smol_csf_oneside}, the corresponding one-sided formulation is:
\begin{equation}
	k_{\text{on}}=\iiint\limits_{\Omega} [\mathbf{n}(\mathbf{x}) \cdot \mathbf{J} (\mathbf{x},t)]  \iiint\limits_{\Omega_a}  [\mathbf{n}(\mathbf{x}) + \mathbf{n}(\mathbf{x}\rq{})] \cdot \nabla_{\mathbf{x}} w(\mathbf{x}-\mathbf{x}', h_r) d\mathbf{x}' d\mathbf{x}.  
	\label{equ:reaction_rate_final}
\end{equation}
If the Robin BC (Eq.~\ref{equ:RobinBC_active}) is enforced, Eq.~\ref{equ:reaction_rate_final} can be reduced to
\begin{equation}
	k_{\text{on}}=\iiint\limits_{\Omega} \alpha p^r(\mathbf{x},t) \iiint\limits_{\Omega_a}  [\mathbf{n}(\mathbf{x}) + \mathbf{n}(\mathbf{x}\rq{})] \cdot \nabla_{\mathbf{x}} w(\mathbf{x}-\mathbf{x}', h_r) d\mathbf{x}' d\mathbf{x} .  
	\label{equ:reaction_rate_final_Robin}
\end{equation}

\end{appendices}
              
\bibliographystyle{bmc-mathphys}
\bibliography{Reference}

%%%%%%%%%%%%%%%%%%%%%%%%%%%%%%%%%%%
%%                               %%
%% Figures                       %%
%%                               %%
%% NB: this is for captions and  %%
%% Titles. All graphics must be  %%
%% submitted separately and NOT  %%
%% included in the Tex document  %%
%%                               %%
%%%%%%%%%%%%%%%%%%%%%%%%%%%%%%%%%%%

%%
%% Do not use \listoffigures as most will included as separate files

\section*{Figures}
\begin{figure}[h!]
	\centering
	\includegraphics[scale=0.6]{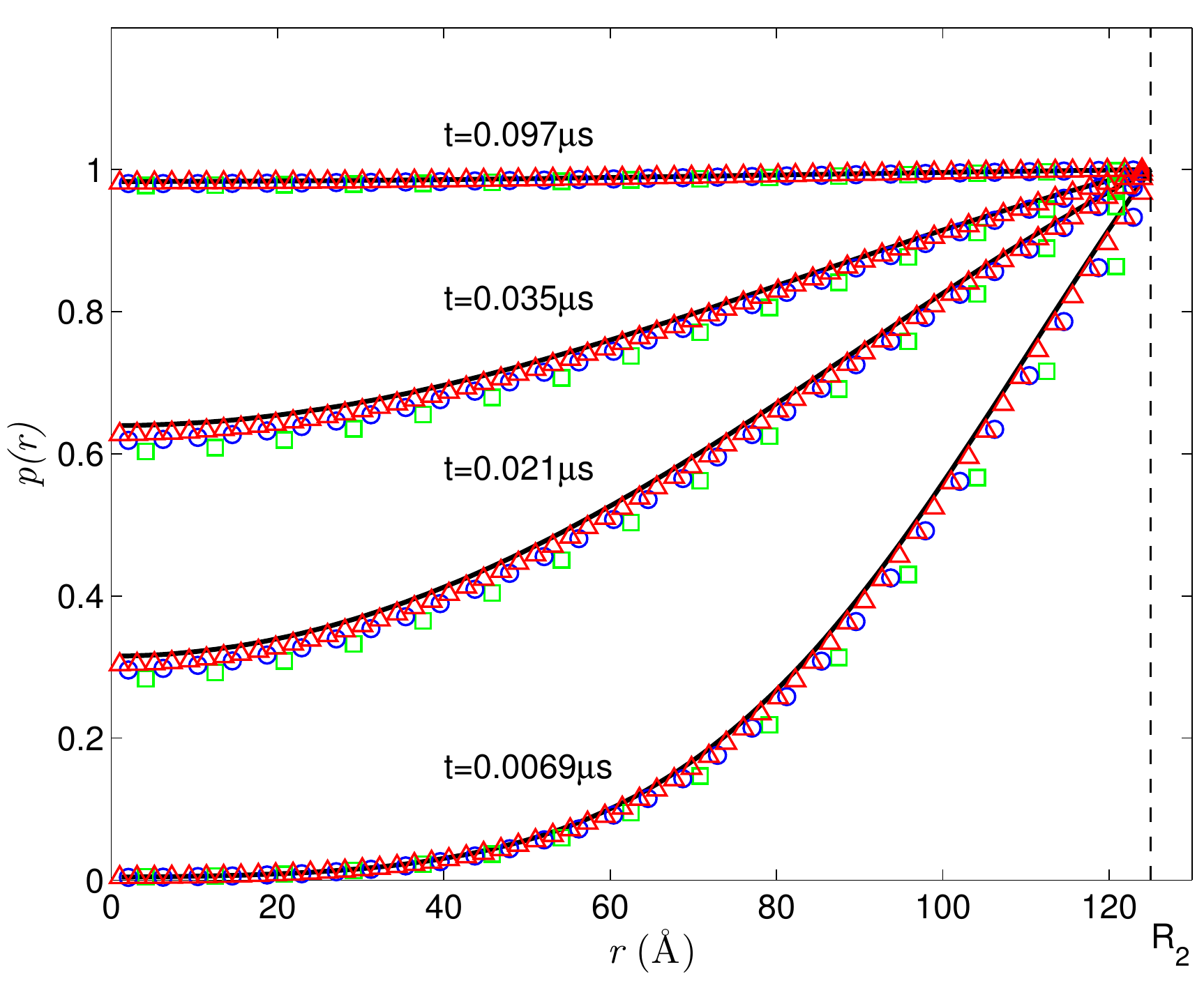}
	\caption{Comparison of SPH solutions to the analytical solution for the Smoluchowski equation subject to the Dirichlet BC on $r=R_2$ at different times with the relative $L_2$ errors for different resolutions. Specifically, $L_2=0.0326$ for $\Delta x=8$\AA (green square), $L_2=0.0180$ for $\Delta x=4$\AA (blue circle), and $L_2=0.0103$ for $\Delta x=2$\AA (red triangle).}
	\label{fig:sphere_dirichlet_time}
\end{figure}

\begin{figure}[h!] 
	\centering
	\includegraphics[scale=0.6]{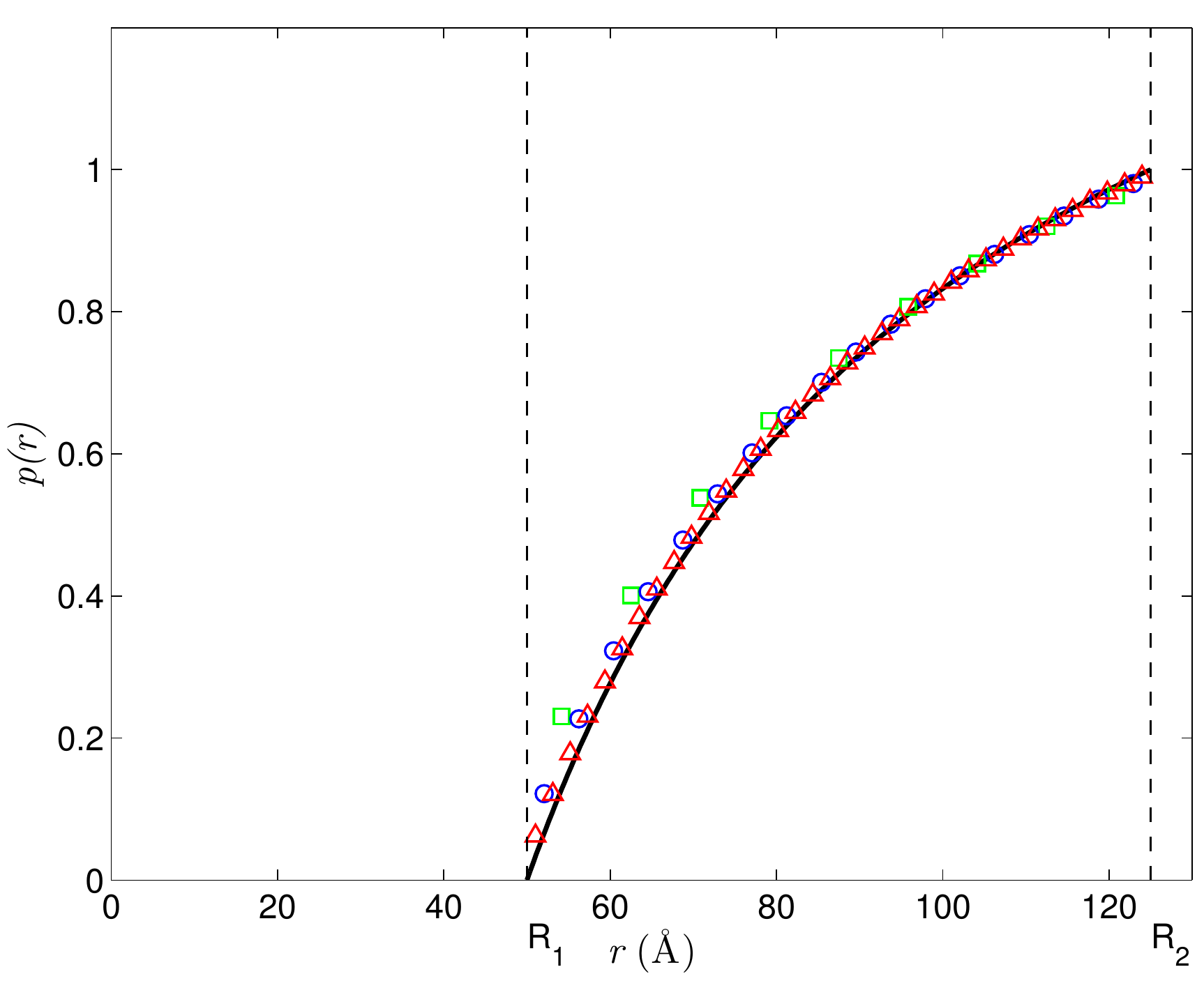}
	\caption{Comparison of SPH solutions to the analytical solution for the Smoluchowski equation subject to the Dirichlet BC on both $r=R_1$ and $r=R_2$ at steady-state with the relative $L_2$ errors for different resolutions. Specifically, $L_2=0.0666$ for $\Delta x=8$\AA (green square), $L_2=0.0321$ for $\Delta x=4$\AA (blue circle), and $L_2=0.0153$ for $\Delta x=2$\AA (red triangle).}
	\label{fig:sphere_dirichlet_steady}
\end{figure}

\begin{figure}[h!]
	\centering
	\includegraphics[scale=0.6]{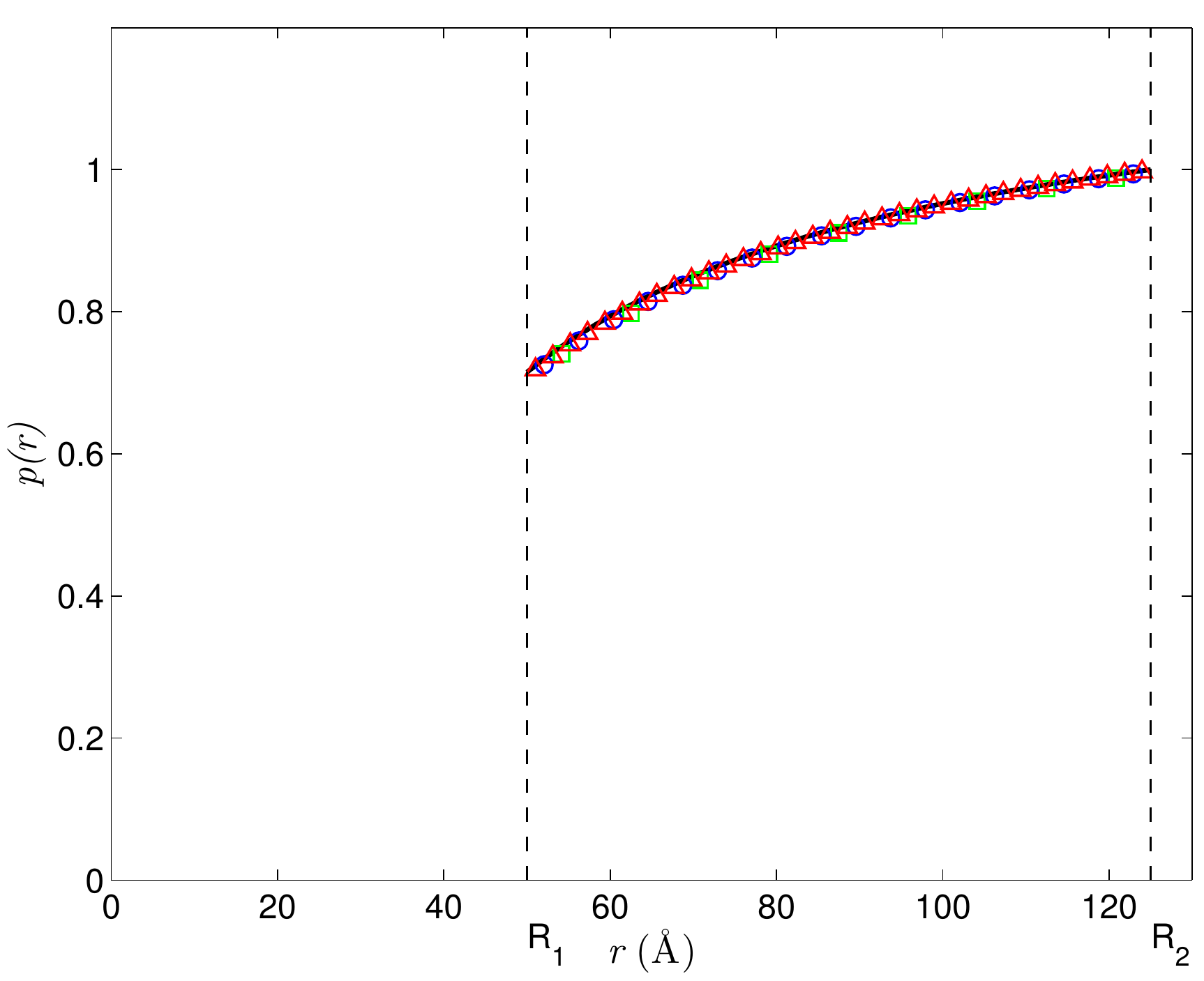}
	\caption{Comparison of SPH solutions (symbol) to the analytical solution (line) for the Smoluchowski equation subject to the Robin BC on $r=R_1$ and Dirichlet BC on $r=R_2$ at steady-state with the relative $L_2$ errors for different resolutions. Specifically, $L_2=0.00914$ for $\Delta x=8$\AA (green square), $L_2=0.00598$ for $\Delta x=4$\AA (blue circle), and $L_2=0.00377$ for $\Delta x=2$\AA (red triangle).}
	\label{fig:sphere_Robin_steady}
\end{figure}

\begin{figure}[h!]
	\centering
	\includegraphics[scale=0.5]{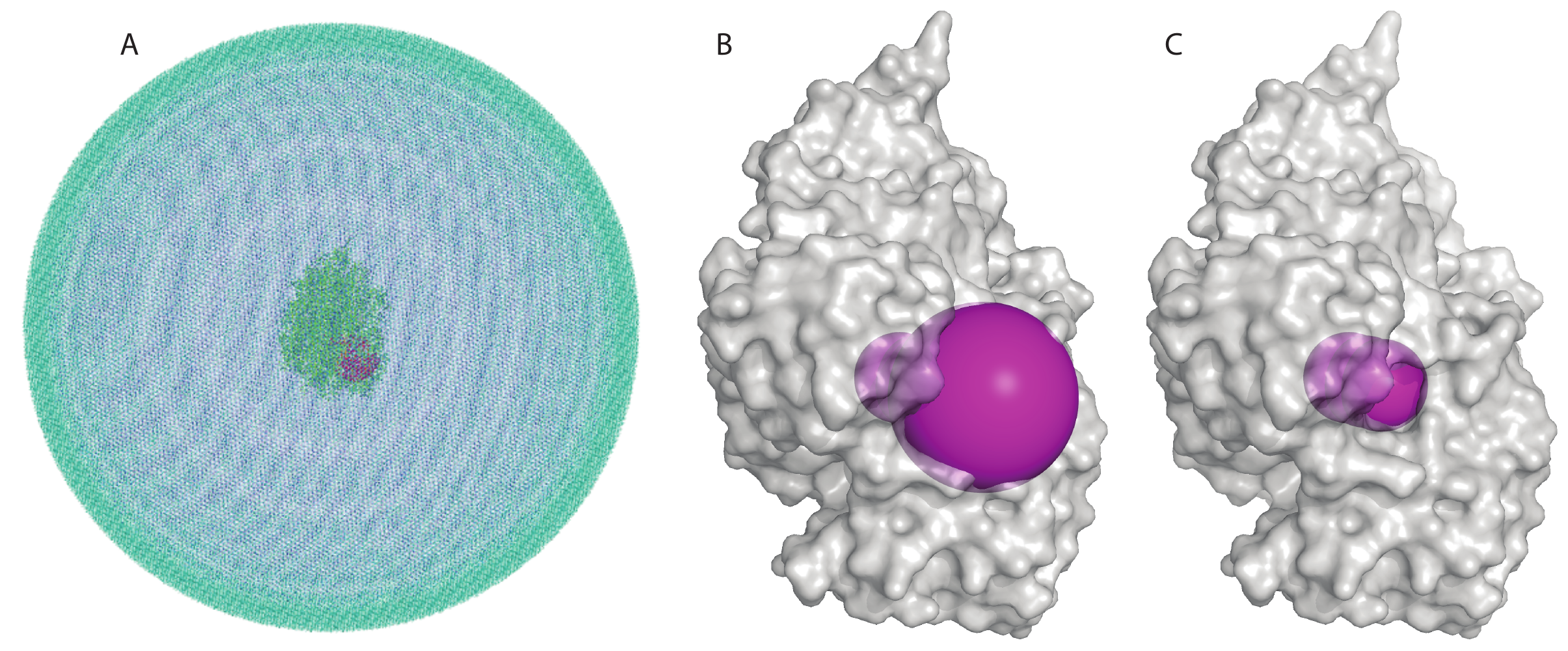}
	\caption{Panel A shows the discretized domain with $R_2 = 128$ {\AA} and the AchE molecule in the center with the reactive boundary shown in purple.  Light blue indicates the outer boundary ($R_2$), blue the solvent, green the protein and magenta the first (outermost) reactive boundary. Panels B and C show reactive boundaries 1 and 4, respectively in magenta spheres.}
	\label{fig:4B82mono_domain}
\end{figure}

\begin{figure}[h!]
	\centering
	\includegraphics[scale=0.5]{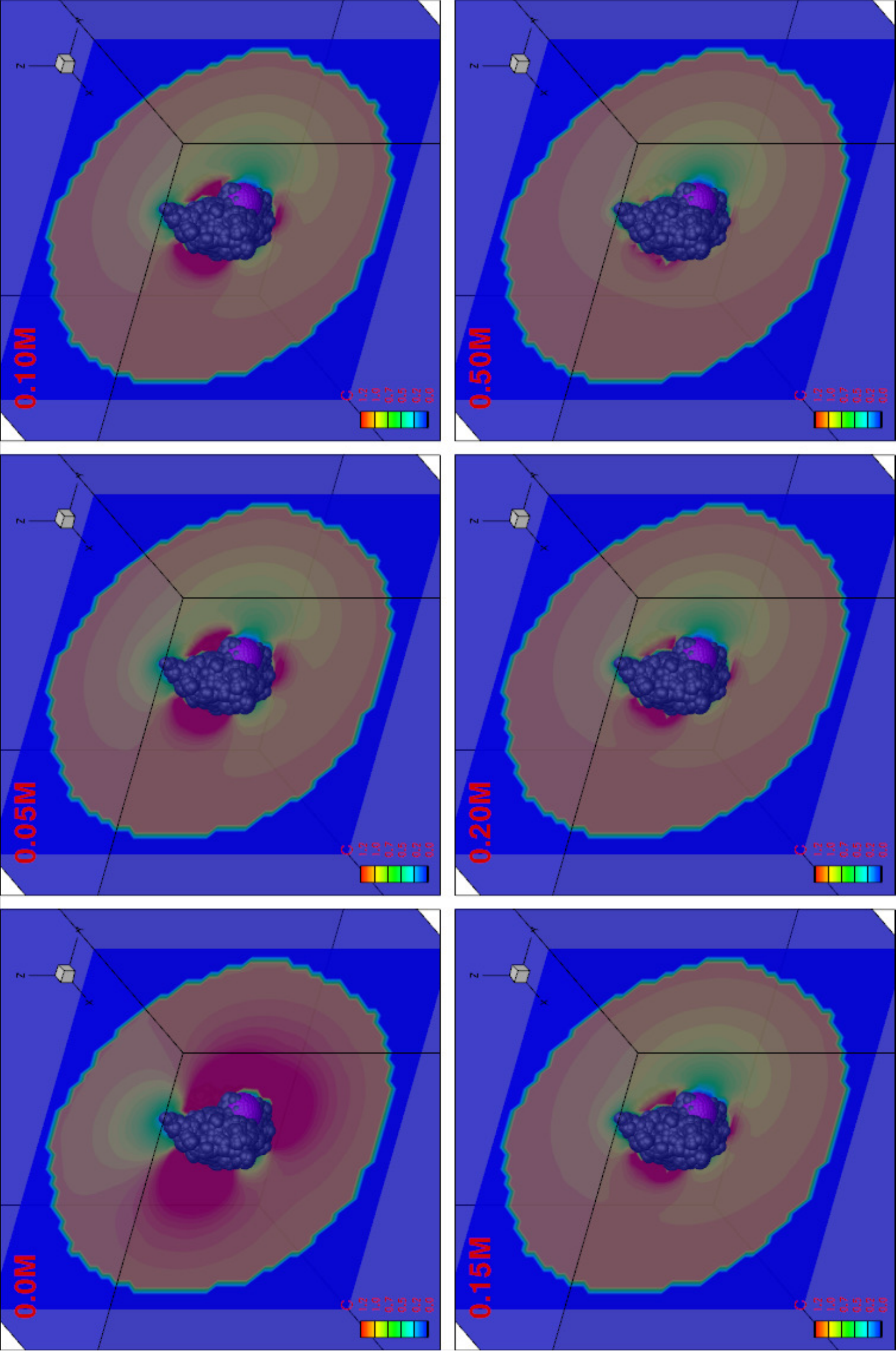}
	\caption{Contour of concentration distribution around mAchE (shown in dark gray) with the Robin BC ($\alpha=8\times10^3$) on reactive boundary 1 at steady state with a range of ionic strengths. Reactive boundary 1 is shown in purple.}
	\label{fig:4B82mono_concen_ionic}
\end{figure}

\begin{figure}[h!]
	\centering
	\includegraphics[scale=0.5,angle=-90]
{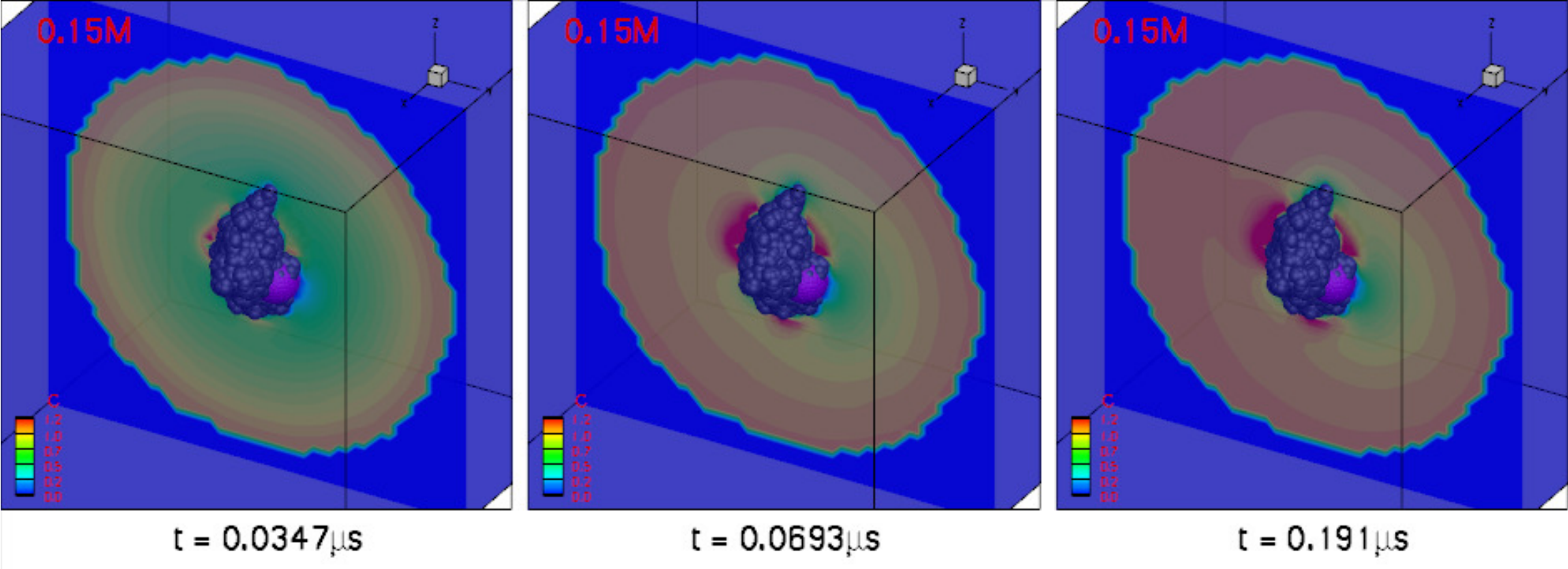}
	\caption{Time evolution of the concentration distribution around mAchE (shown in gray) with a Robin BC ($\alpha=8\times10^3$) on reactive boundary 1 at 0.15 M ionic strength. Reactive boundary 1 is shown in purple.}
	\label{fig:4B82mono_concen_015M_time}
\end{figure}

\begin{figure}[h!]
	\centering
	\includegraphics[scale=0.5]{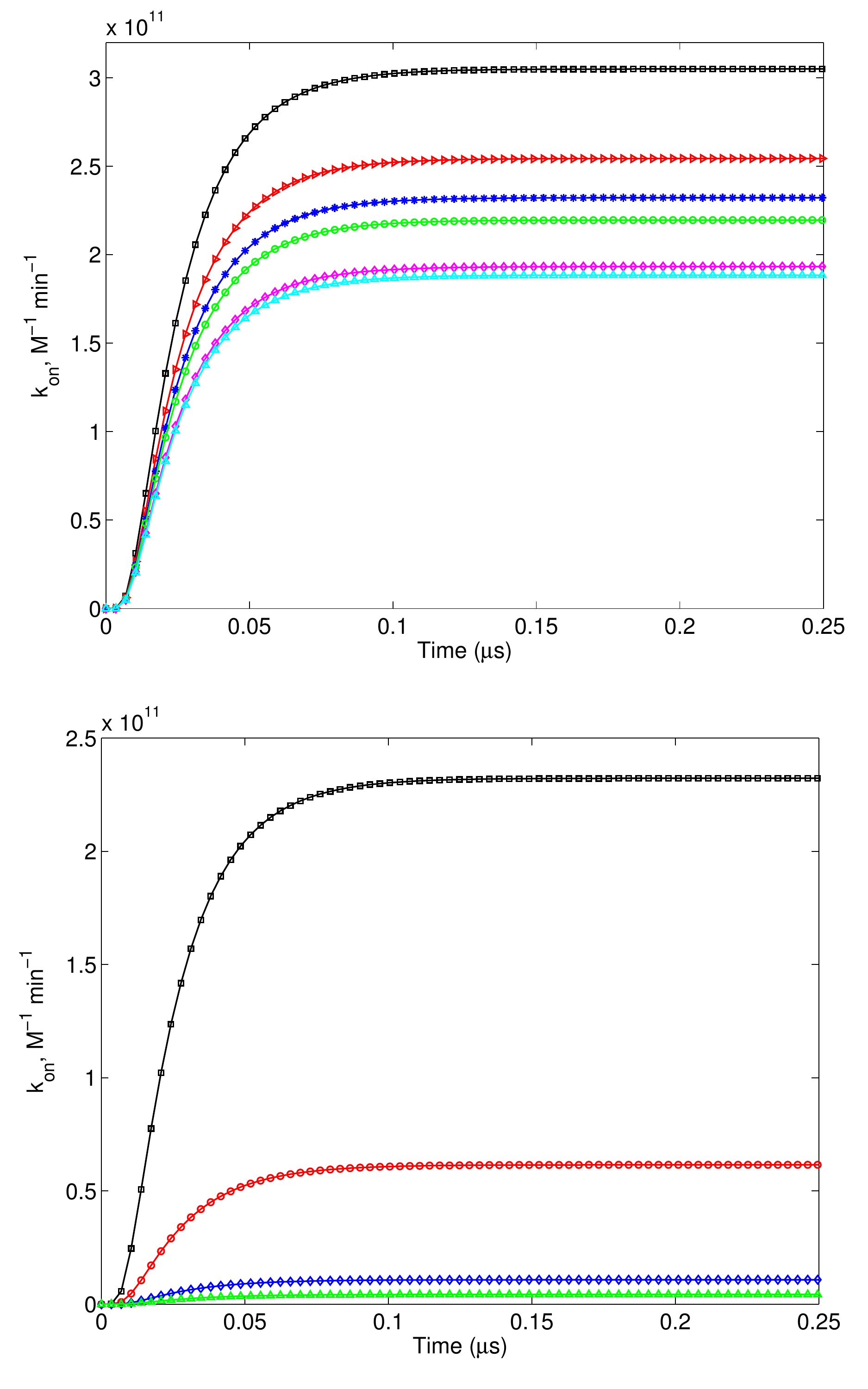}
	\caption{({Left}) $k_\mathrm{on}$ as a function of $t$ on reactive boundary 1 at different ionic strengths.  Black square: 0.05M; red right-pointing triangle: 0.10M; blue asterisk: 0.15M; green circle: 0.20M; magenta diamond: 0.50M; cyan triangle: 0.67M. ({Right}) $k_\mathrm{on}$ as a function of $t$ on reactive boundaries 1-4, respectively, at 0.15 M ionic strength. Black square: reactive boundary 1; red circle: reactive boundary 2; blue diamond: reactive boundary 3; green triangle: reactive boundary 4.}
	\label{fig:4B82mono_k_ionic}
\end{figure}

\begin{figure}[h!]
	\centering
	\includegraphics[scale=0.5]{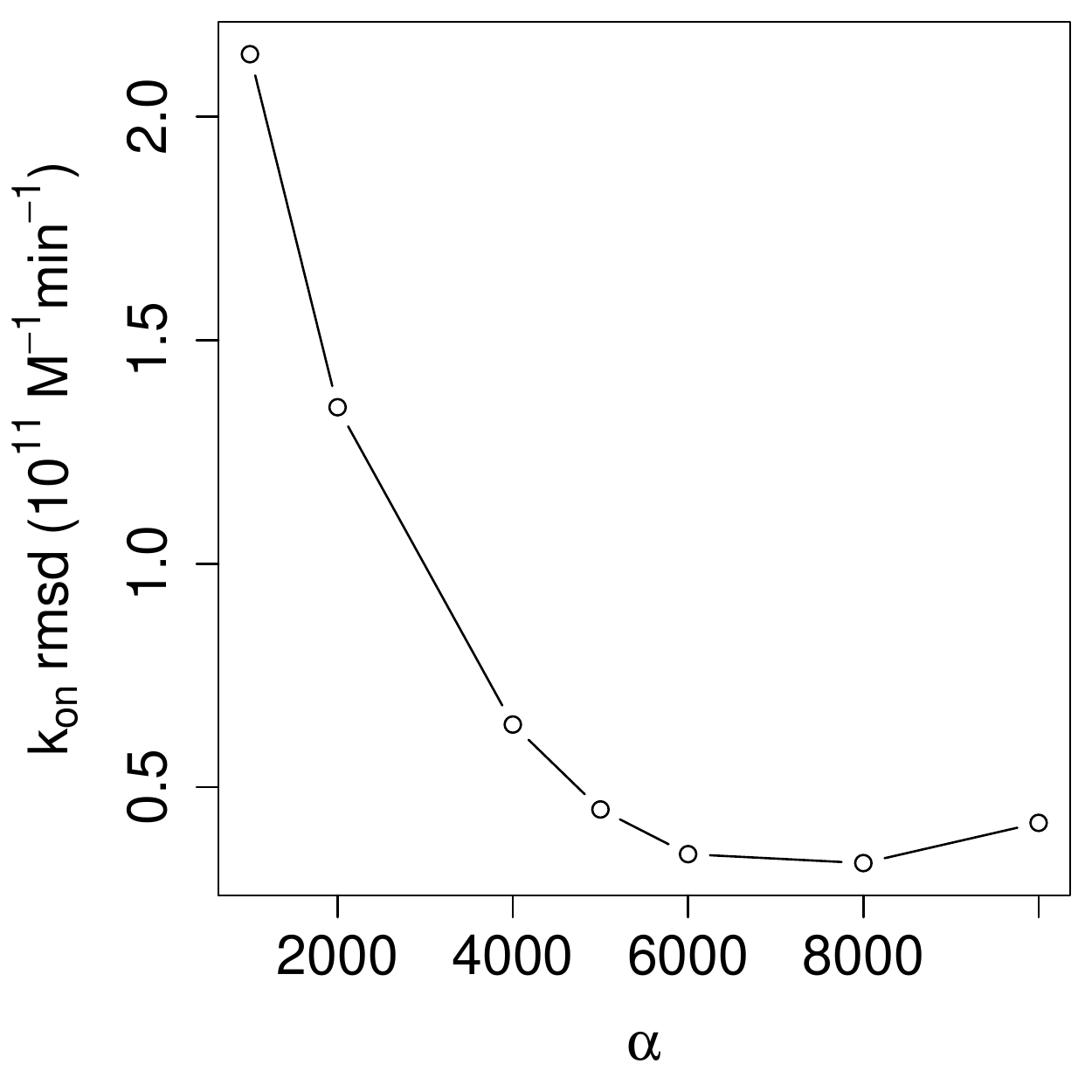}
	\caption{Root mean square deviation (RMSD) of computed by SPH to experimental reaction rates (over 0-0.67 M ionic strengths) vs. $\alpha$ for the Robin BC.}
	\label{fig:4B82mono_k_alpha}
\end{figure}

\begin{figure}[h!]
	\centering
	\includegraphics[scale=0.5]{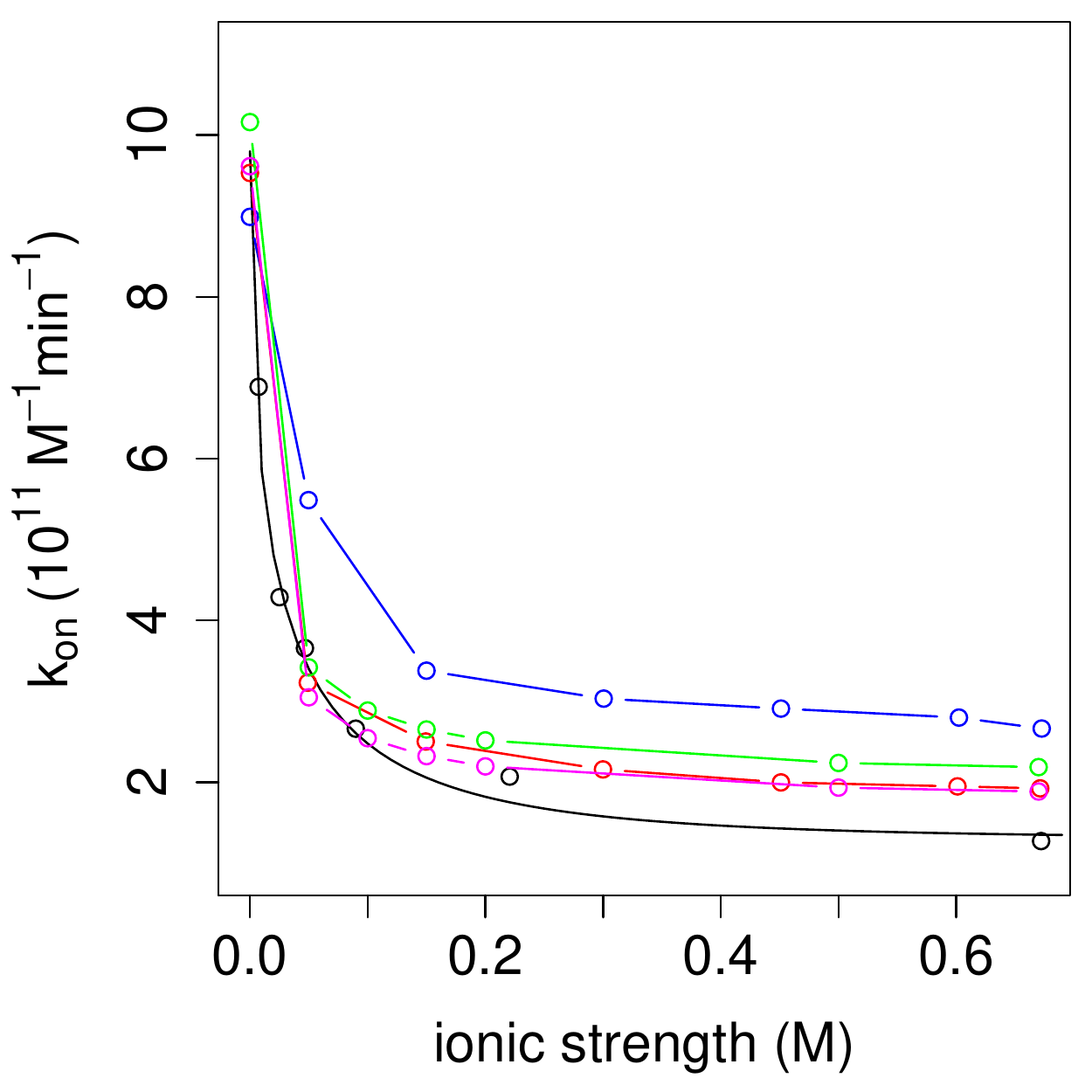}
	\caption{Reaction rates of mAChE on reactive boundary 1 obtained from different methods.  Black: from experimental data \cite{Radic1997} (symbol) and fitted (line) to the Debye-H\"uckel limiting law (Eq.~\ref{equ:Debye_Huckel}); blue: from BD \cite{Tara1998}; red: from FEM with Dirichlet BC \cite{SongFEM2004}; green:  from SPH with Dirichlet BC; magenta: from SPH with Robin BC using $\alpha=8\times10^3$.  For standardization, both computed and experimental data are fitted to the Debye-H\"uckel limiting law.}
	\label{fig:4B82mono_k_datacomp}
\end{figure}

\begin{figure}[h!]
	\centering
	\includegraphics[scale=0.7]{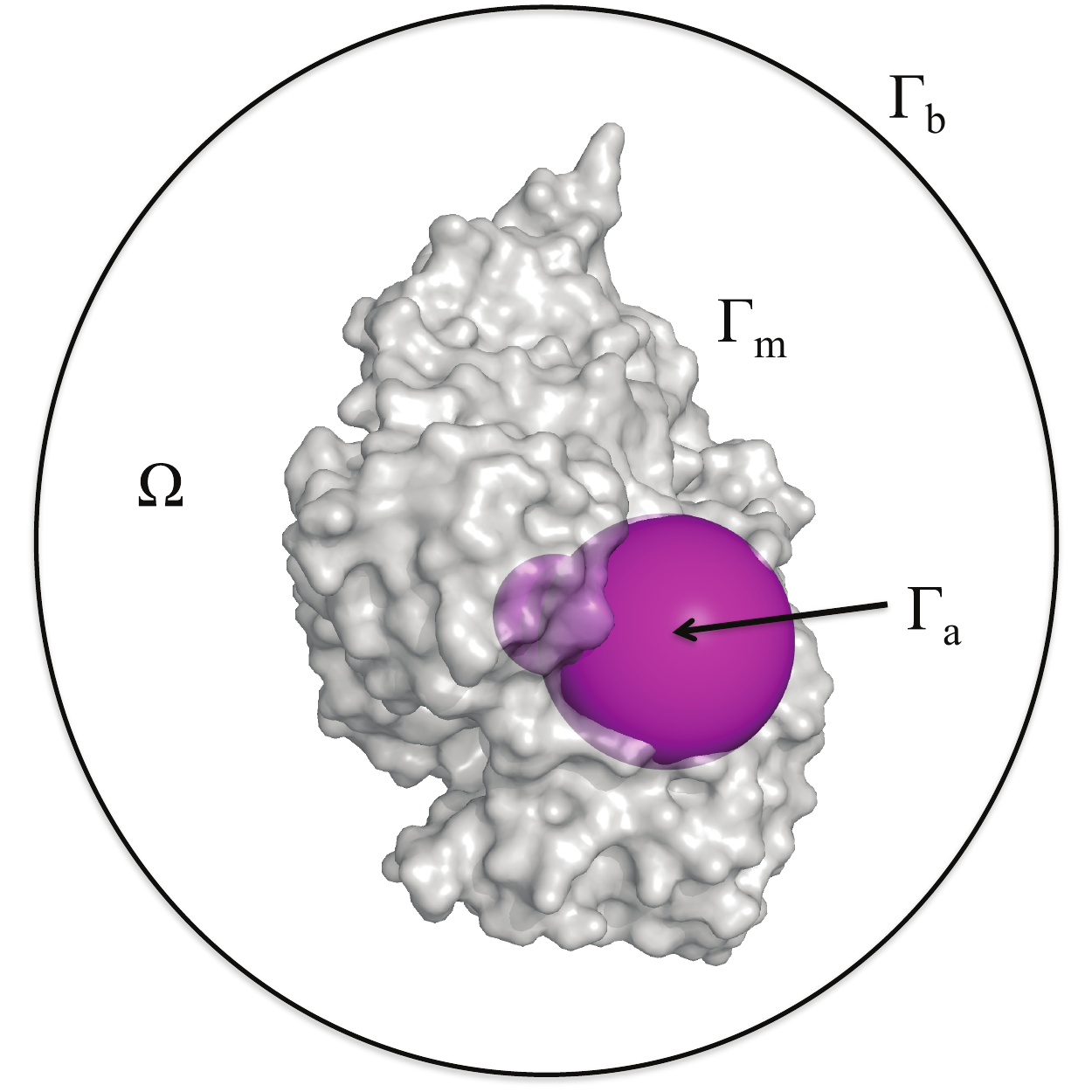}
	\caption{Illustration of the simulation domain and all boundaries: $\Gamma_b$ indicates the outer boundary, $\Gamma_m$ the molecular surface, and $\Gamma_a$ the reactive boundary 1; $\Omega$ indicates the problem domain between $\Gamma_b$ and $\Gamma_a\cup \Gamma_m$.}
	\label{fig:solution_domain}
\end{figure}

%%%%%%%%%%%%%%%%%%%%%%%%%%%%%%%%%%%
%%                               %%
%% Tables                        %%
%%                               %%
%%%%%%%%%%%%%%%%%%%%%%%%%%%%%%%%%%%

%% Use of \listoftables is discouraged.
%%
\section*{Tables}
\begin{table}[t]
	\caption{Comparison of Debye-H\"uckel fits vs.\ ionic strength between experiment and simulations. RMSD of data to experimental $k_\mathrm{on}$ is calculated over the range of ionic strengths between 0 and 0.67 M. The unit of $k_\mathrm{on}^0$, $k_\mathrm{on}^\mathrm{H}$ and RMSD is $10^{11}\mathrm{M^{-1} {min}^{-1}}$. And the error is the standard deviation of parameter fits using nonlinear least squares.}
	\begin{center}
		\begin{tabular}{c  c  c  c  c  c  c}
			\hline
			Data & $k_\mathrm{on}^0$  &  error &  $k_\mathrm{on}^\mathrm{H}$ & $Z_E$ & error & RMSD \\
			\hline
			Radic et al \cite{Radic1997} & 9.80 &	0.60 & 1.30 & 2.30 &	0.20 & 0.00\\
			BD &	9.08 & 0.30 & 2.66 &	1.67 & 0.15 & 1.52\\
			FEM & 9.52 &	0.11 & 1.92 & 2.79 &	0.11 & 0.37\\
			SPH (Dirichlet) & 10.15 &	0.09 &	2.19 &	2.92 &	0.08 &	0.57\\
			SPH (Robin,  $\alpha=8\times10^3$) &	9.61	& 0.08 &	1.88 &	2.96 &	0.08 &	0.33\\
			hexa-mutant (experiment) & 1.80 & 0.10 & 0.57 & 1.20	& 0.20 & 0.00\\
			hexa-mutant (SPH, Robin,  $\alpha=8\times10^3$) & 2.23	& 0.00 &	1.77 &	2.37 &	0.02 &	0.88\\
			hexa-mutant (SPH, Robin,  $\alpha=6\times10^3$) & 2.01	& 0.00 &	1.58 &	2.33 &	0.02 &	0.70\\
			\hline
		\end{tabular}
	\end{center}
	\label{tab:4B82mono_k_datacomp}
\end{table}

\end{backmatter}

\end{document}